\documentclass[useAMS,usenatbib]{mn2e}

\usepackage{txfonts,amssymb}
\usepackage{amsfonts}
\usepackage{epsfig}

\newlength{\colwidth}
\newcommand{\spose}[1]{\hbox to 0pt{#1\hss}}
\newcommand{\simlt}{\mathrel{\spose{\lower 3pt\hbox{$\mathchar"218$}}
     \raise 2.0pt\hbox{$\mathchar"13C$}}}
\newcommand{\simgt}{\mathrel{\spose{\lower 3pt\hbox{$\mathchar"218$}}
     \raise 2.0pt\hbox{$\mathchar"13E$}}}

\title[The ratio of $L_{\rm acc}$ to $L_\ast$
in
T~Tauri Stars] {Disc evolution and the relationship between $L_{\rm acc}$ and
$L_\ast$ in T~Tauri stars}

\author[I. Tilling, C. J. Clarke,  J. E. Pringle \& C. A. Tout]
{I. Tilling, C. J. Clarke, J. E. Pringle and C. A. Tout 
\\
Institute of Astronomy, The Observatories, Madingley Road, Cambridge, CB3 0HA \\} 

\begin{document}
\maketitle

\begin{abstract}

We investigate the evolution of accretion luminosity $L_{\rm acc}$ and
stellar luminosity ${L_\ast}$ in pre-mainsequence stars. We make the
assumption that when the star appears as a Class II object, the major
phase of accretion is long past, and the accretion disc has entered its
asymptotic phase. We use an approximate stellar evolution scheme for
accreting pre-mainsequence stars based on Hartmann, Cassen \& Kenyon,
1997. We show that the observed range of values $k = L_{\rm
  acc}/L_\ast$ between 0.01 and 1 can be reproduced if the values of
the disc mass fraction $M_{\rm disc}/M_*$ at the start of the T~Tauri
phase lie in the range 0.01 -- 0.2, independent of stellar mass. 
We also show that the observed upper bound of $L_{\rm acc} \sim
L_\ast$ is a generic feature of such disc accretion. We conclude that
as long as the data uniformly fills the region between this
upper bound and observational detection thresholds, then
the degeneracies between age, mass and accretion history severely
limit the use of this data for constraining possible scalings between
disc properties and stellar mass. 

\end{abstract}


\section{Introduction}

The claimed relationship between accretion rate, $\dot{M}$, and mass,
$M_*$, in pre-mainsequence stars (e.g. Natta, Testi \& Randich 2006
and references therein) originates from the more direct observational
result that in Class II pre-mainsequence stars the accretion
luminosity, $L_{\rm acc}$, is similarly found to correlate with
stellar luminosity, $L_\ast$. The accretion luminosity is deduced from
the luminosity in the emission lines of hydrogen (in Natta et al.,
2006, these are Pa$\beta$ and Br$\gamma$), together with modeling of
the emission process (Natta et al. 2004; Calvet et al. 2004). The
modeling appears to be relatively robust. In order to deduce the
quantities $\dot{M}$ and $M_*$ from the quantities $L_{\rm acc}$ and
$L_\ast$, it is necessary to be able to deduce the stellar properties.
This is done by placing the object in a Hertzsprung-Russell diagram
and making use of theoretical pre-mainsequence tracks for
non-accreting stars (e.g. D'Antona \& Mazzitelli 1997).

The result of
this exercise indicated a correlation between
$\dot{M}$ and $M_*$ of the form $\dot{M} \propto M_*^{\alpha}$, where
$\alpha$ is close to $2$. The simplest theoretical expectation
(in the case of disc accretion where neither the fraction of material
in the disc nor the disc's viscous timescale scale systematically
with stellar mass) is instead that $\dot{M} \propto M_*$. The claimed  steeper
than linear relationship thus motivated a variety of theoretical ideas
about specific scalings of disc parameters with stellar mass
(Alexander \& Armitage 2006, Dullemond et al 2006). 

\begin{figure}
\resizebox{\colwidth}{!}{\includegraphics[angle=0]{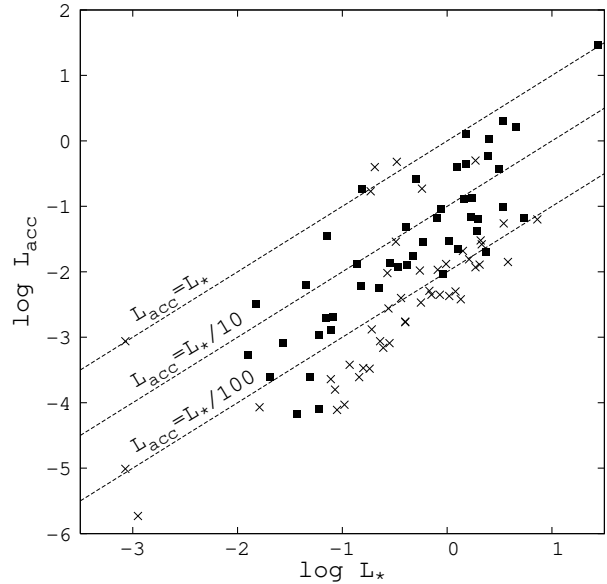}}
\caption{The distribution of Classical T Tauri stars in 
Ophiuchus in the $L_\ast$, $L_{\rm acc}$ plane. Detections (filled squares)
and upper limits (crosses) deduced from emission line data.
Adapted from Natta et al 2006.} \label{hereifany2}
\end{figure}

Clarke \& Pringle (2006) however suggested that the claimed correlation
between $\dot{M}$ and $M_*$ could be understood as follows.
They
noted that the distribution of datapoints in the plane of
$L_\ast$ versus $L_{\rm acc}$ (Figure 1) more or less fills a region that is
bounded, at high $L_{\rm acc}$ by the condition $k = L_{\rm acc}/L_\ast \sim
1$ and, at low $L_{\rm acc}$,  by observational detection thresholds, which roughly
follow a relation of the form $L_{\rm acc} \propto L_\ast^{1.6}$. 
They noted that when this relation between $L_{\rm acc}$ and $ L_\ast$ is combined with specific
assumptions about the relationship between $L_\ast$ and $M_*$, and between $L_\ast$ and stellar radius
$R_\ast$ (based on 
placing the stars on pre-mainsequence tracks at a particular age,
or narrow spread of ages) then the resulting relationship
between $\dot{M}$ and $M_*$ is indeed $\dot{M} \propto M_*^2$. They thus
claimed that it is currently impossible to reject 
the possibility that
the claimed, steeper than linear, relationship is an artefact of
detection biases.

Clarke \& Pringle (2006) also pointed out that 
the distribution of detections in the 
$L_{\rm acc}$ -- $L_\ast$ plane may nevertheless  be used to tell us something
about conventional accretion disc evolutionary models. In this paper
we explore the extent to which this might be achieved. The first
question which needs to be addressed is why there is a spread of
values of $L_{\rm acc}$ at a given value of $L_\ast$. 
Secondly we need to understand {\it why} the upper locus of
detections corresponds to $L_{\rm acc} \sim L_\ast$, since there
seems no obvious
reason why this should correspond to a detection
threshold or selection effect.

 We begin here  (Section 2) by setting out a simple argument why -- as a generic
property of accretion from a disc with mass much less than the
stellar mass -- one would
expect the distribution to be bounded by the condition 
$L_{\rm acc} \sim L_\ast$. In Section 3 we use the ideas of Hartmann, Cassen \& Kenyon (1997) to
develop an approximate evolutionary code for accreting
pre-mainsequence stars and demonstrate its applicability by comparison
with the tracks of pre-mainsequence stars computed by Tout, Livio \&
Bonnell (1999). In Section 4 we apply the code to
pre-mainsequence stars accreting from an accretion disc with a
declining accretion rate and investigate the range of disc parameters
required to obtain the observed range of $L_{\rm acc}/L_\ast$. We draw
our conclusions in Section 5.

\section {The maximum ratio of accretion luminosity to stellar luminosity}

From a theoretical point of view there are three relevant timescales
in the case of star gaining mass by accretion from a disc.
The first is

\begin{equation}
t_M = M_*/\dot{M},
\end{equation}
which is the current timescale on which the stellar mass is
increasing. The second is

\begin{equation}
t_{\rm disc} = M_{\rm disc}/\dot{M},
\end{equation}
which is the current timescale on which the disc mass is
decreasing. The third is the Kelvin-Helmholtz timescale

\begin{equation}
t_{\rm KH} = \frac{GM_*^2}{R_\ast L_\ast},
\end{equation}
which is the timescale on which the star can radiate its thermal
energy. It is also the timescale on which a star can come into thermal
equilibrium. On the pre-main sequence, when the luminosity of the star
is mainly due to its contraction under gravity, $t_{\rm KH}$ is also
the evolutionary timescale.

We see immediately  (since $L_{\rm acc} \sim G M_* \dot M/R$) that  

\begin{equation}
\frac{L_{\rm acc}}{L_\ast} = \frac{t_{\rm KH}}{t_M}.
\end{equation}

The pre-mainsequence stars in which we are interested, Class~II (the
Classical T~Tauri stars), all have the following properties: 

(i) They have passed through the major phase of mass accretion which
occurs during the embedded states Class 0 and Class I.  This implies
typically that they have been evolving for some time with their own
gravitational energy as the major energy source. Deuterium burning
interrupts this briefly but only delays contraction by a factor of 2
or so.  Thus for these stars we may expect that they have an age
$T \approx t_{\rm KH}$ roughly.

(ii) The masses of their accretion discs are now small, $M_{\rm disc}
< M_*$. Further, accretion disc models at such late stages -- i.e. the so
called asymptotic stage when most
of the mass that was originally contained in them has been accreted
on to the central object --  exhibit a  power law decline in $\dot {M}$
with time and thus have  the generic property that the age of the
disc is about $t_{\rm disc}$. If most of the star's lifetime has
been spent accreting in this asymptotic regime,  
we may roughly equate the age of the disc and the age
of the star and write $T \approx t_{\rm disc}$.

Given this, we see immediately that, for these stars, we expect $t_M =
M_*/\dot{M} > M_{\rm disc}/\dot{M} = t_{\rm disc} \approx t_{\rm KH}$ and
therefore that $L_{\rm acc} <  L_\ast$.  

In the following section we test this argument by evolving a suite
of model star-disc systems in the $L_\ast - L_{\rm acc}$ plane.

\section{Stellar evolution calculations} 

Rather than attempting to carry out full stellar evolution
computations of accreting and evolving pre-mainsequence stars (e.g. Tout
et al. 1999), we here adopt the simplified approach of Hartmann,
Cassen \& Kenyon (1997).

We begin with a stellar core of mass $M_{\rm i} = 0.1\,\rm M_\odot$ and
radius $R_{\rm i} = 3 \,\rm R_\odot$. We prescribe the accretion rate
$\dot{M}(t)$ and thus the stellar mass as a function of time

\begin{equation}
  M_*(t) = M_{\rm i} + \int_0^t \dot{M}(t) dt. 
\end{equation}

Hartmann et al. (1997) write the energy equation as

\begin{equation}
L_* = -{{3}\over{7}} {{GM_*^2}\over {R}} \biggl[{{1}\over{3}} {{\dot M}
    \over {M_*}} + {{\dot R}\over{R}}\biggl] \,  + \, L_D.
\end{equation}
In deriving this equation they assume that the star is always in
hydrostatic balance and, being fully convective, can be treated as an
$n=3/2$ polytrope.  Account is taken of the conversion of
gravitational energy into thermal energy via $pdV$ work as well energy
input from deuterium burning $L_D$ and energy loss via radiative
cooling at the photosphere $L_*$. In the form written above, it is
assumed that the material accreting on to the star arrives with zero
thermal energy.  This is probably a good approximation to the case
where material enters the star via a radiative magnetospheric shock or
from a disc boundary layer.  The rate of energy input to the star
delivered by deuterium burning is parameterised (following the
expression given by Stahler 1988 for an $n=3/2$ polytrope) by

\begin{equation}
  L_D/L_\odot = 1.92 \times 10^{17} XfX_{\rm D}
  \left({{M_*}\over{M_\odot}}\right)^{13.8}
  \left({{R}\over{R_\odot}}\right)^{-14.8},
\end{equation}
where $X$ is the mass fraction of hydrogen and $f$ is the fraction of
deuterium remaining in the star relative to the accreted material.  We
assume that the accreted material has the same abundance as the star
before deuterium begins burning and we take this initial mass fraction
to be $X_{\rm D} = 3.5\times 10^{-5}$ as did Tout et al. (1999). We
fix the hydrogen mass fraction at $X = 0.7$.\footnote{Our notation
  differs from that of Stahler (1988) and Hartmann et al. (1997)
  because we prefer not to use their non-standard definition of [D/H].
  We recover their $\beta_{\rm D}\equiv Q_{\rm D}X_{\rm D}$ in
  equation~(\ref{deut}) with $X_{\rm D} = 3.5\times 10^{-5}$.}

The variation of $f$ is given by (equivalent to Stahler 1988).

\begin{equation}
\label{deut}
 {{df}\over{dt}} = {{\dot M}\over{M_*}}\left(1-f-{{L_D}\over{Q_{\rm
D}X_{\rm D}
     \dot M}}\right),
\end{equation}
where $Q_{\rm D} = 2.63 \times 10^{18}\,\rm erg\,g^{-1}$ is the energy available
from fusion of deuterium.  Because of timestep problems we
set $f=0$ once the value of $f$ falls below some small value $f_{\rm
min}$. Except where explicitly mentioned, we take $f_{\rm min} =
10^{-6}$. 

For a fully convective star, the stellar entropy is controlled by the
outer (photospheric) boundary condition which takes account of the
strong temperature sensitivity of the surface opacity for stars on the
Hayashi track. Hartmann et al. (1997) adopt a fit to the
(non-accreting) stellar evolutionary tracks of D'Antona \& Mazzitelli
(1994) in the form

\begin{equation}
\label{photobc}
 L_*/L_\odot = \biggl({{M_*}\over{0.5 M_\odot}}\biggr)^{0.9}
 \biggl({{R}\over {2 R_\odot}}\biggr)^{2.34}.
\end{equation}
They stress that this fit is only approximate and is adequate only in
the mass range  $0.3 \le M_*/{\rm M_\odot} \le 1$.

\begin{figure}
\resizebox{\colwidth}{!}{\includegraphics[angle=0]{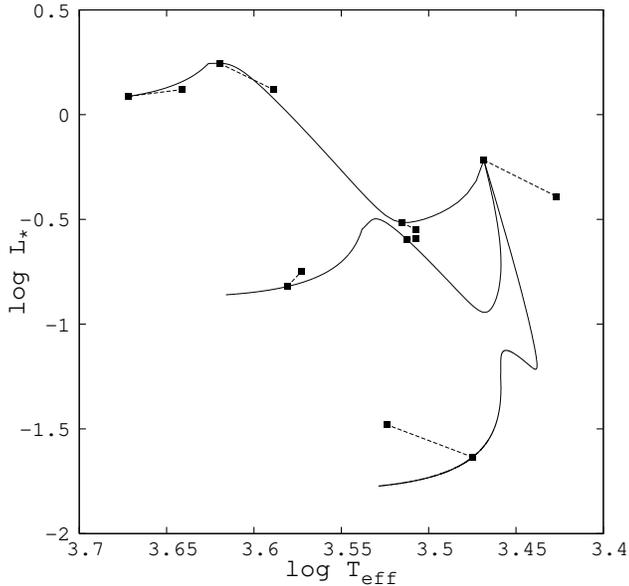}}
\caption{Accreting pre-mainsequence tracks computed using the
simplified evolutionary scheme, with $f_{\rm min} = 0.001$. Tracks
correspond to constant accretion rates of (from top to bottom)
$10^{-6}$, $10^{-7}$ and $10^{-8}\,\rm M_\odot\,yr^{-1}$.  Initially
the mass is $0.1\,\rm M_\odot$ and the radius is $3\,\rm R_\odot$. The
tracks are terminated when the approximations cease to be valid 
(because a radiative core has developed or hydrogen burning
begins in the core) or at a maximum age of $20$ Myr.  Points are plotted along the tracks at masses of
0.1, 0.2, 0.5~and $1.0\,\rm M_\odot$ up to the maximum mass allowed by
each track. The corresponding points from the Tout et al. (1999)
tracks are joined to these by dotted lines.  } \label{hereifany1}
\end{figure}

In order to provide some estimate of the accuracy of this simplified
evolutionary scheme, we compared the results against the full stellar
evolution calculations of Tout et al. (1999), for the case of constant
accretion rate (having first satisfied ourselves that stars in the
appropriate mass range in Tout et al
are indeed well fit by the parameterisation of the  
photospheric condition given by (9)).
The results are illustrated in Fig.~2, for accretion rates
$\dot{M} = 10^{-6}, 10^{-7}$ and~$10^{-8}\,\rm M_\odot\,yr^{-1}$. On each track
(where possible) we plot the point at which the stellar mass is
$M_*/M_\odot = 0.2, 0.5$ and~$1.0$ together with the corresponding points on
the tracks. Both sets of computations had the same initial
values of $M_* = 0.1\,\rm M_\odot$ and $R = 3\,\rm R_\odot$. The stars first
descend a Hayashi track, while being driven across to higher
effective temperature by the accreting matter.  When the core temperature reaches about $10^6\,$K they evolve
up the deuterium-burning main sequence.  Once deuterium fuel is exhausted they resume their descent
towards the zero-age main
sequence, leaving their Hayashi tracks when the stellar luminosity drops
to the point that the Kelvin timescale becomes greater than the accretion
timescale. At this point, accretion is again important and the tracks
are then driven to higher effective temperatures.
Up to these points our simplified model appears to be in reasonably good agreement with Tout et al (1999)'s models,
even at early times and low masses, when the equations are formally
incorrect.  All the points are accurate to within a factor of
two in luminosity and to within 10 per cent in effective temperature. We
conclude that this simplified model is adequate for the task in hand.

\section{Application to T~Tauri stars}

In view of our ignorance about formation history of pre-mainsequence
stars in terms of their accretion history, we need to make some
assumptions about the form of $\dot{M}(t)$. In doing so, we are
guided by some simple physical ideas. It needs to be borne in mind,
however, that although our simple assumptions give rise to results
which provide a reasonable description of the data, this does not imply
that the assumptions are necessarily correct nor that they are the
only ones which might work.

We assume that most of the mass which has accreted on to a T
Tauri star has been processed through an accretion disc. This
assumption, coupled with the observational fact that protostellar
discs are at most 10 -- 20 per cent of the stellar mass and usually
much less, tells us that the evolution of the accretion disc has
progressed into its late, asymptotic phase. Lynden-Bell \& Pringle
(1974; see also Hartmann et al. 1988) present similarity solutions
for accretion disc evolution using the simple assumption that the
kinematic viscosity varies as a power law of radius. They find that
for such discs, the disc accretion rate varies at late times as a
power law in time with $\dot{M} \propto t^{-\eta}$, for some $\eta >
1$.  This is true more generally for discs in which the viscosity also
varies as powers of local disc properties, such as surface density
(Pringle 1991).  Here we take $\eta = 1.5$,
corresponding to kinematic viscosity being proportional to radius
(Hartmann et al. 1988).

We assume that T~Tauri stars only become visible at a time $t =
t_0$ after the onset of formation and take $t_0 = 5 \times 10^5\,$yr.
We consider four sets of models in terms of the disc mass
fraction $f_{\rm disc} = M_{\rm disc}/M_{\rm f}$ at time $t_0$, where
$M_{\rm f}$ is the final stellar mass. We consider $f_{\rm disc}
= 0.2, 0.1, 0.037$ and $0.014$. In order to stay within the
credibility limits of our evolutionary scheme we
consider final masses only in the range $0.4 \le M_{\rm f}/M_\odot \le 1.0$.

We also need to define the stellar accretion rate from time $t=0$. At
$t=0$ we start with $M_* = M_{\rm i} = 0.1\,\rm M_\odot$. In principle, the
evolution of the star for $t > t_0$ can depend critically on the
accretion rate prior to this point. This is simply because the
evolution timescale for a pre-mainsequence star is always comparable
to its age, so that it has never had time to forget its initial
conditions. In order to minimise this effect, we assume that the
accretion rate does not increase with time. This is probably
reasonable for stars which form in isolation (e.g. Shu, Adam \&
Lizano 1987), but may not be valid for more dynamical modes of star
formation, when late interaction and merging may be normal
(e.g. Bate, Bonnell \& Bromm 2002). To be specific we assume
that the accretion rate is of the form
\begin{equation}
\dot{M} = \dot{M}_{\rm i} = {\rm const.},  \; 0 \le t \le t_{\rm s}
\end{equation}
and that
\begin{equation}
\dot{M} = \dot{M}_{\rm i} \left( \frac{t}{t_{\rm s}} \right)^{-\eta},
\; t_{\rm s} \le t,
\end{equation}
where $t_{\rm s}$ and $\dot{M}_{\rm i}$ are chosen to ensure that the integral
of $\dot{M}$ between zero and infinity is $M_{\rm f} - M_{\rm i}$ and that
the integral of $\dot{M}$ from $t_0$ to infinity is $f_{\rm disc} M_*$
(see Table~1). We note that, since $t_0 \gg t_{\rm s}$, the disc is
always in the asymptotic regime at the stage that it  is classified as a
T Tauri star. We stress that the form of the evolutionary prescription that 
we
have adopted for $\dot M(t)$ prior to $t_0$ is not the only one that produces
a given disc and stellar mass at $t_0$ and that  - depending on the temporal
history and location of matter infall 
onto the disc -    
a variety of prior histories, with the same integrated mass, are possible.
Our intention here is however not to explore all possibilities but
to adopt a plausible and easily adjustable prescription with which
one can investigate the system's evolution in the $L_{acc}-L_*$ plane.  

\begin{table*}
\centering
\begin{minipage}{\colwidth}
\caption{The parameters for the runs shown in Fig.~3. $M_{\rm f}$
  is the final stellar mass. $f_{\rm disc}$ is the fraction of
  $M_{\rm f}$ which is still in the disc at time $t_0 = 5 \times
  10^5\,$yr. The accretion rate takes on a constant value of
  $\dot{M}_{\rm i}$ until time $t_{\rm s}$ and thereafter declines as
  $\propto t^{-\eta}$ with $\eta = 1.5$ \label{ifany}}
{\begin{tabular}{llll}
$M_{\rm f}/M_\odot$ & $f_{\rm disc}$ & $t_{\rm s}/$yr & $\dot{M}_{\rm i}/\rm M_\odot\,yr^{-1}$ \\ 
\hline
1.0&0.2&$5.5 \times 10^4$&$5.4 \times 10^{-6}$ \\
1.0&0.1&$1.4 \times 10^4$&$2.2 \times 10^{-5}$ \\
1.0&0.037&$1.9 \times 10^3$&$1.5 \times 10^{-4}$ \\
1.0&0.014&$2.8 \times 10^2$&$1.1 \times 10^{-3}$ \\
\hline
0.4&0.2&$8.0 \times 10^4$&$1.2 \times 10^{-6}$ \\
0.4&0.1&$2.0 \times 10^4$&$5.0 \times 10^{-6}$ \\
0.4&0.037&$2.8 \times 10^3$&$3.6 \times 10^{-5}$ \\
0.4&0.014&$4.0 \times 10^2$&$2.5 \times 10^{-4}$ \\
\hline
\end{tabular}}
\end{minipage}
\end{table*}

\begin{figure}
\resizebox{\colwidth}{!}{\includegraphics[angle=0]{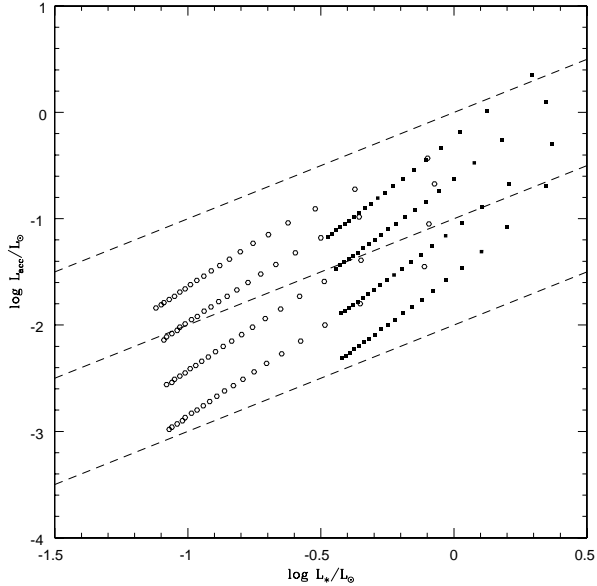}}
\caption{Accretion luminosity $L_{\rm acc}$ against
stellar luminosity $L_\ast$ for the various runs with parameters given
in Table~1. The filled symbols lead to final mass $M_{\rm f} =
1.0\,\rm M_\odot$ and the open symbols to $M_{\rm f} = 0.4\,\rm
M_\odot$. Each line of symbols represents one evolutionary sequence,
starting at top right at time $t_0 = 5 \times 10^5\,$yr and finishing
bottom left at time $t = 10^7\,$yr, with one symbol plotted every $5
\times 10^5\,$yr. For each mass, the disc mass fractions at time $t_0$
are $f_{\rm disc} = 0.2$ (uppermost track) $= 0.1, = 0.037$ and $=
0.014$ (lowermost track). The dashed diagonal lines represent $L_{\rm
  acc}/L_\odot= 1.0, 0.1$ and~$0.01$ respectively.  } \label{hereifany}
\end{figure}

In Fig.~3 we show the results of these computations showing the
accretion luminosity ($L_{\rm acc}$) as a function of stellar
luminosity ($L_\ast$). We plot results for two final masses $M_{\rm
  f}/M_\odot = 1.0$ and $0.4$ and for the four initial disc mass
fractions $f_{\rm disc} = 0.014, 0.037, 0.1$ and $0.2$. We plot the
positions starting at time $t = 5 \times 10^5\,$yr and then at
intervals of $5 \times 10^5\,$yr until time $t = 10^7\,$yr. In Table 1
we give the values of $t_{\rm s}$ and of $\dot{M}_{\rm i}$
corresponding to these parameters. 
We note that in all our models
$t_{\rm s} \ll t_0$; thus, by the time that the star is deemed
to be a Classical T Tauri star, the accretion rate has been
declining as a power law over the majority of the system's lifetime. 

From Fig.~3 we see that our data points more or less fill the range
$10^{-2} < L_{\rm acc}/L_\ast < 1$ in the range of values of $L_\ast$
which are accessible to our computations. It is reasonable to expect that
with a more detailed evolutionary scheme, valid at lower masses, this
trend would continue to lower values of $L_\ast$. Thus these simple
assumptions, coupled with this simplified evolutionary scheme are
capable of producing the typical observed spread of points in the
$L_{\rm acc} - L_\ast$ diagram (see Fig. ~1).

The stars in Fig.~3 are evolving at essentially constant mass because,
by assumption, $M_{\rm disc} \ll M_*$ and have already consumed
all of their initial deuterium. At constant mass, Equation~\ref{photobc}
gives $L_\ast \propto R^{2.34}$. For a star evolving on the
Kelvin-Helmholtz timescale this implies $R_\ast \propto t^{-1/3.34}$
and thus that $L_\ast \propto t^{-0.7}$. Since $L_{\rm acc} =
GM_* \dot{M}/R_\ast$ and we have assumed $\dot{M} \propto t^{-
\eta}$ we have $L_{\rm acc} \propto t^{-(\eta - 1/3.34)}$. From
this we deduce that as time varies, we have approximately
\begin{equation}
L_{\rm acc} \propto L_\ast^{(\eta - 0.3)/0.7}.
\end{equation}
For $\eta = 1.5$ this is $L_{\rm acc} \propto
L_\ast^{1.7}$. Thus the slope of the tracks in Fig.~3 is a result of
the ratio of the rates at which the star and the accretion disc
evolve.  Owing to the photospheric boundary condition this ratio just depends on $\eta$.

  It might be thought suggestive that the relation $L_{acc} \propto L_\ast^{!.7}$ (which holds in the case $\eta = 1.5$, at least over the range of $L_\ast$
for which (9) is valid), is very close to the mean relationship in the
observational data. It is then tempting to see this as observational support
for a viscosity law with this value of the power law index. However, we stress 
that when we experimented with a different viscosity law, which changed
the slope of the individual stellar trajectories according to equation
(12), the distribution of stars in the $L_\ast -L_{\rm acc}$ plane
was indistinguishable, filling, as it did, the available parameter space
between $L_{\rm acc} \sim L_\ast$ and observational detection thresholds.

We note that at fixed $L_\ast$ there is a trend for lower values of
$L_{\rm acc}$ to correspond to lower values of $M_{\rm disc}$, the
disc mass fraction at the time when the star is deemed to appear as a
classical T~Tauri star. Thus in this picture the spread of values of
$L_{\rm acc}/L_\ast$ at fixed $L_\ast$ comes about because of the
spread of disc masses at late times. Such a spread could
be due to a number of reasons such as a range of initial
disc sizes (i.e. initial angular momenta) or a variety of disc
truncations caused by dynamical interactions.

There is also a trend for lower values of stellar mass $M_*$ to
correspond to lower values of $L_\ast$. It is evident, however, that
the tracks in this diagram intermingle to some extent.  At the
times plotted $M_{\rm disc} \ll M_*$, so each track corresponds essentially
to a fixed stellar mass. Thus it is clear that at a given value of
$L_\ast$ there can be a spread of stellar masses depending on the
assumed stellar age.

\section{Conclusions}

We have shown, using simplified stellar evolution calculations for
stars subject to a time dependent accretion history, that a plausible
morphology of the $L_* -  L_{\rm acc}$ plane for T~Tauri stars has the
following features,

 (i) an upper locus corresponding to $L_* \approx L_{\rm acc}$ and

 (ii) diagonal tracks for falling accretion rates as stars descend Hayashi
tracks.

We have argued that (i) is a generic feature of stars which are
accreting from a disc reservoir with mass less than the central star
coupled with the assumption that the mass depletion timescale of the 
disc is comparable to
the age  of the star. \footnote {We may in fact argue that the
existence of such an upper limit in the data is a good signature of 
temporally declining accretion through a disc, since if Classical T Tauri stars
instead accreted external gas at a constant rate (e.g. Padoan  et al 2005), no
such feature would be apparent in the data}.
The first is necessary for stars to be classified
as Classical T~Tauri stars while the second results from the gradual
decay of the accretion rate.

We show that the slope of of the diagonal tracks (ii) is related to
the index of the power law decline of accretion rate with time at late
times.  This itself depends on the scaling of viscosity with radius in
the disc. For reference, if $\nu \propto R$, then $L_{\rm acc} \propto
L_\ast^{1.7}$ along such tracks, whereas if $\nu \propto R^{1.5}$
then
$L_{\rm acc} \propto L_\ast^{2.4}$.  
  
 We however argue that, as long as observational datapoints appear
to fill the whole region of the $L_\ast - L_{\rm acc}$ plane that is bounded
by the upper limit (i) and luminosity dependent sensitivity thresholds,
it will remain impossible to deduce what  viscosity law or what 
dependence of disc parameters on stellar mass will be  required
in order to populate this region. This is because of the severe
degeneracies that exist between stellar mass, age and accretion history
(our results show that in order to populate the desired region, it
is necessary that there is, at the least,  a spread in both stellar mass and 
initial disc properties). It is not, however,   necessary to
posit any particular scaling of  disc properties with
stellar mass in order to populate the plane, in contrast to the
hypotheses of Alexander \& Armitage 2006 and Dullemond et al
2006. Likewise, although,
as noted above, the trajectories followed by individual stars in this
plane are a sensitive diagnostic of the disc viscosity law, this information
is washed out in the case of an ensemble of stars which simply fill
this  plane. 

 We therefore conclude that such data could only yield useful
insights if regions of the plane turned out to be unpopulated
(or very sparsely populated). There is a shadow of a suggestion
in Figure 1 that at low $L_\ast$  the data falls further
below the locus $L_{\rm acc} = L_\ast$ than at higher luminosities
although, given the sample size, this difference is not statistically
significant (Clarke \& Pringle 2006). Any such edge in the distribution 
could in principle re-open the possibility of constraining disc models.
We however emphasise that this low luminosity regime can in any case
not be explored by the simple evolutionary models that we employ here,
and that -- if it becomes necessary to investigate this regime --  it will
be essential to use full stellar evolution calculations.

  Finally, we note that we have chosen to relate our study to observational
data in the $L_{\rm acc} -L_*$ plane because the ratio of these
quantities is straightforwardly derivable from emission line
equivalent widths. Thus this comparison is more direct than the alternative,
i.e. the  comparison  with observational data that has been transformed into the
plane of $\dot M$ versus $M_*$ via the use of isochrone fitting and
the application of theoretical mass-radius relations. It is nevertheless
worth  noting 
that the upper locus of $L_{acc} \sim L_*$ corresponds to a line with
$\dot M \propto M_*$. Although it is often taken as self-evident that
this is the expected ratio (in the case that the maximum
initial ratio of disc mass to stellar mass is independent of stellar mass),
this actually only corresponds to $\dot M \propto M_*$ in the
case that the disc's initial viscous time is independent of stellar
mass. Given the possible variety of viscosity values and radii
of protostellar discs it is not at all evident that these should
conspire together to give the same viscous timescale --  until, that is,
that one realises that all discs in the asymptotic regime have
viscous timescale which is equal to the disc age. The conclusion
that the upper locus should follow the relation $\dot M \propto M_*$
is thus a very general one.

 In summary, we conclude, as in the study of  Clarke \& Pringle
(2006), that the observed population of the $L_{\rm acc} - L_*$ plane,
and the consequent relationship between $\dot {M}$ and $M_*$, is strongly
driven by luminosity dependent sensitivity thresholds and an upper
locus at $L_{\rm acc} \approx L_\ast$.  The main new ingredient
injected by our numerical calculations and plausibility arguments is
that we now understand that this observed upper locus can be
understood as a consequence of plausible accretion histories, given
the current properties of the relevant pre-mainsequence stars. In
addition, our computations indicate the range of initial disc properties
that are required in order to populate the region of parameter space
occupied by observational datapoints:
 we are able to reproduce the
observed range of values of $L_{\rm acc}/L_\ast$ at a given value of
$L_\ast$ by assuming once the T~Tauri phase has been reached, the disc
mass fractions lie roughly in the range $0.01 \le M_{\rm disc}/M_* \le
0.2$, independent of stellar mass.

\section*{Acknowledgments}

CAT thanks Churchill College for his Fellowship.
We acknowledge useful discussions with Lee Hartmann and Kees Dullemond.

\end{document}